\documentclass[twocolumn,showpacs,preprintnumbers,amsmath,amssymb]{revtex4}

\usepackage{graphicx}
\usepackage{dcolumn}
\usepackage{bm}

\begin{document}

\title{Bose-Einstein Condensates with Cavity-Mediated Spin-Orbit Coupling}
\author{Y. Deng$^{1}$, J. Cheng$^{2}$, H. Jing$^{3}$, and S. Yi$^{1}$}
\affiliation{$^1$State Key Laboratory of Theoretical Physics,
Institute of Theoretical Physics, Chinese Academy of Sciences, P.O.
Box 2735, Beijing 100190, China}
\affiliation{$^{2}$Department of
Physics, South China University of Technology, Guangzhou 510640,
China}
\affiliation{$^{3}$Department of Physics, Henan Normal
University, Xinxiang 453007, China}

\date{\today}

\begin{abstract}
We propose a novel scheme to generate the spin-orbit coupling for a condensate placed inside an optical cavity by using a standing wave and a traveling wave. It is shown that the interplay of the laser lights and the cavity gives rise to rich quantum phases. Our scheme also generates an nontrivial lattice of the magnetic flux, which may facilitate the study of the exotic quantum phases. 
\end{abstract}

\maketitle

{\em Introduction}.---The recent experimental realizations of the spin-orbit (SO) coupling in ultracold atomic
gases~\cite{Lin11, Wang12, Cheuk12, Zhang12} offer the unique opportunity for studying a wide range of many-body physics in a highly controlled way~\cite{Dalibard11,Galitski13}. To date, the theoretical investigations have covered the quantum phases of SO coupled Bose-Einstein condensates (BECs)~\cite{Sinha11, Wu11, Wang10, Ho11, Xu11,Hu12,Deng12,Wilson} and exotic superfluid orders of ultracold Fermi gases~\cite{Sato09, Jiang11,Gong11, Hu11, Zhai11, gongm, Radic12, zhangyi, Hu13}. Of particular interest, the quantum anomalous Hall effect and chiral topological superfluid phase with two-dimensional SO interaction were predicted with Raman laser configurations~\cite{xiong}. Meanwhile, the capability of preparing BECs in optical cavities opens a new avenue for exploring the strongly correlated phenomena in atomic gases~\cite{Esslinger10,Mottl,Dimer07,Maschler08,Nagy08,Domokos10,Keeling10,
Baumann11, Gopalakrishnan11,Strack11, Habibian13}. When loaded into an optical cavity, all atoms in the BEC couple to the same cavity field which induces long-range interactions between atoms. This cavity-mediated long-range interaction leads to the observations of the self-organized supersolid phase~\cite{Esslinger10}, the roton-type mode softening~\cite{Mottl}, and the quantum spin-glass phase~\cite{Gopalakrishnan11, Strack11, Habibian13}. Moreover, the inherent leakage of the cavity to monitor the phase transition with nondemolition measurement the cavity leakage~\cite{Esslinger10, Mottl,Baumann11}.

Motivated by these developments, we investigate, in this Letter, the quantum phases of a SO coupled condensate in an optical cavity. We propose to generate SO coupling using Raman lights for a condensate placed inside an optical cavity. We show that, depending on the strengths of the pumping field and the Raman coupling, the system admits the checkerboard and stripe phases. More remarkably, the cavity-induced Raman coupling can gives rise to a quantum phase characterized by a lattice of vortex-antivortex pairs (LVAP). In this phase, the light fields induce an optical flux lattice for the addressed spin state~\cite{cooper}, which allows for the studies of the quantum anomalous Hall effect and the topological superfluids~\cite{xiong}. While the optical flux lattices can be realized with other schemes, our method requires a simpler the laser configuration. We also show that the cavity-induced self-organization also emerges in the checkerboard and LVAP phases.

\begin{figure}[tbp]
\includegraphics[width=0.9\columnwidth]{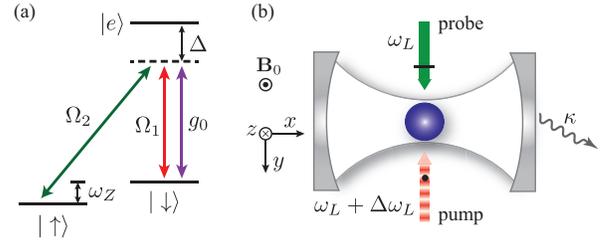}
\caption{(color online). (a) Level diagram. (b) Scheme for creating spin-orbit coupling inside an optical cavity.}\label{model}
\end{figure}

{\em Model}.---We consider a gas of $N$ bosonic $\Lambda$-type atoms placed inside a high-finesse cavity. Figure~\ref{model}(a) illustrates the level structure of the atom. The atomic transition frequency between the electronic ground states (denoted as $|\uparrow\rangle$ and $|\downarrow\rangle$) and the excited state (denoted as $|e\rangle$) is $\omega_{a}$. The magnetic quantum numbers of the electronic states satisfy the conditions: $m_{|\uparrow\rangle}-m_{|\downarrow\rangle}=1$ and $m_{|e\rangle}=m_{|\downarrow\rangle}$. As shown in Fig.~\ref{model}(b), a bias magnetic field ${\mathbf B}_{0}$ is applied along the negative $z$-axis, which defines the quantization axis and produces a Zeeman shift $\hbar\omega_Z$ between $|\uparrow\rangle$ and $|\downarrow\rangle$. Moreover, atoms are illuminated along the $y$-axis by a standing-wave pump laser with frequency $\omega_L+\Delta\omega_L$ and a plane-wave probe laser with frequency $\omega_L$, which are linearly polarized along the $z$- and $x$-axes, respectively. The atom-pump detuning is defined as $\Delta = \omega_a -(\omega_L + \Delta \omega_L)$. As can be seen, the atomic transition $|\downarrow\rangle\leftrightarrow|e\rangle$ ($|\uparrow\rangle\leftrightarrow|e\rangle$) is driven by the pump (probe) laser; the corresponding Rabi frequency is $\Omega_1\cos k_1y$ ($\Omega_2 e^{ik_2 y}$) with $k_1$ ($k_2$) being the wave vector. 

As to the cavity, its mode frequency, wave vector, and decay rate of the cavity are $\omega_{c}$, $k$, and $\kappa$, respectively. In general, the cavity supports both $\pi$- and $\sigma^-$-polarized photon fields originated from the collective Bragg scattering. However, under the conditions of the pump-cavity resonance ($\omega_c = \omega_L+\Delta \omega_L$) and the large Zeeman shift ($|\omega_Z/\kappa|\gg 1$), the Bragg scattering into the $\sigma^-$-polarized cavity mode can be suppressed~\cite{Wilk07}. As a result, only the $\pi$-polarized cavity field generated by the vacuum-stimulated Raman transition is relevant in our system. Therefore, the atomic states $|\downarrow\rangle$ and $|e\rangle$ is also coupled by the cavity field with a coupling strength $g_{0}$. 

In the large atom-pump detuning limit, $|\Omega_{1,2}/\Delta|\ll1$ and $|g_{0}/\Delta|\ll1$, the excited atomic state can be adiabatically eliminated which leads to the effective many-body Hamiltonian for the pseudospin-$1/2$ atoms
\begin{align}
\hat H=&\sum_{\sigma\sigma'}\int d{\mathbf r}\hat\psi_{\sigma}^{\dag}({\mathbf r})[\hat h_{\sigma\sigma'}+V_{\rm ext}({\mathbf r})\delta_{\sigma\sigma'}]\hat\psi_{\sigma'}({\mathbf r})\nonumber\\
&+\frac{1}{2}\sum_{\sigma\sigma'}\frac{4\pi\hbar^{2}a_{\sigma\sigma'}}{m}\!\!\int \!d{\mathbf r}\hat\psi_{\sigma}^{\dag}({\mathbf r})\hat\psi_{\sigma'}^{\dag}({\mathbf r})\hat\psi_{\sigma'}({\mathbf r})\hat\psi_{\sigma}({\mathbf r}),\label{manyh}
\end{align}
where $m$ is the mass of the atom, $\hat {\boldsymbol h}$ is the single-particle Hamiltonian (see below), $\hat\psi_{\sigma=\uparrow,\downarrow}$ is the field operator for spin-$\sigma$ atom, the external trap $V_{\rm ext}({\mathbf r})=m\left[\nu_{\perp}^{2}(x^{2}+y^{2})+\nu_{z}^{2}z^{2}\right]/2$ is assumed to be spin independent, and $a_{\sigma\sigma'}$ are the $s$-wave scattering lengths between the spin-$\sigma$ and -$\sigma'$ atoms. The single-particle Hamiltonian takes the form
\begin{eqnarray}
\hat {\boldsymbol h} =\frac{{\mathbf p}^{2}}{2m}\hat {\boldsymbol I}+  \hbar\left(\!
\begin{array}{cc}
-\frac{\delta}{2} & \hat M_{-}(x,y)  \\
\hat M_{-}^{\dag}(x,y) & \frac{\delta}{2}+\hat M_{z}(x,y)
\end{array}
\!\right)\!,\;\label{singleh}
\end{eqnarray}
where $\delta = \omega_Z + \Delta \omega_L+\Omega_2^2/\Delta$ is the effective two-photon detuning, $\hat M_{-}=\left(-\Omega \cos ky + \Omega_c \hat{a} \cos kx\right)e^{-iky}$ with $\hat a$ being the annihilation operator of the cavity photon, $\Omega = -\Omega_1\Omega_2/\Delta$ and $\Omega_c = g_0 \Omega_2/\Delta$ being the effective Raman coupling strengths due to, respectively, the classical and cavity fields. Moreover, $\hat M_{z}=U_{1}\cos^{2} ky + \eta(\hat{a}^\dag\! + \hat{a})\cos kx\cos ky + U_{0}\hat{a}^{\dag}\hat{a}\cos^{2} kx$, where $\eta =-g_0\Omega_1/\Delta$ is the maximum scattering rate, and $U_0 = -g_0^2/\Delta$ and $U_{1} = -{\Omega^2_1}/{\Delta}$ are the optical Stark shifts. Here, we have taken $k_{2}\approx k_{1}(=k)$, which is valid because $|\Delta\omega_{L}/\omega_{L}|\ll1$ is generally fulfilled. As can be seen, $\hat {\boldsymbol h}$ contains two SO coupling terms in $\hat M_{-}^{\dag}$, corresponding to the Raman transitions induced, respectively, by the classical and cavity fields. We remark that the proposed model is experimentally realizable, in particular, with Cr and Dy atoms~\cite{Deng12}.

The cavity field $\hat a$ in $\hat H$ can be eliminated by using the dynamical equations for the field operators: $i\hbar\partial\hat a/\partial t=[\hat a,\hat H]$ and $i\hbar\partial\hat\psi_{\sigma}({\mathbf r})/\partial t=[\hat\psi_{\sigma}({\mathbf r}),\hat H]$. For large decay rate, $\kappa\gg |\eta|,|\Omega_c|$, the cavity field reaches a steady-state on a time scale much faster than the external atomic motion~\cite{Esslinger10, Mottl, Maschler08, Nagy08}. As a result, one may set $\partial{\hat a}/\partial t=0$, which leads to the formal solution for the steady cavity field. Upon substituting $\hat a$ back into the dynamical equations of $\hat\psi_{\sigma}$, one can derive an effective Hamiltonian for the atomic operators. 

To proceed further, we adopt the mean-field approximation by replacing $\hat\psi_{\sigma}$ in their dynamical equations with the condensate wave functions $\psi_{\sigma}({\mathbf r})=\langle\hat\psi_{\sigma}({\mathbf r})\rangle$, i.e.,
\begin{equation}
i\hbar\frac{\partial\psi_{\sigma}}{\partial t}=\sum_{\sigma'}\left(\hat h_{\sigma\sigma'}+V_{\rm ext}\delta_{\sigma\sigma'}+\frac{4\pi\hbar^{2}a_{\sigma\sigma'}}{m}\psi_{\sigma'}^{*}\psi_{\sigma}\right)\psi_{\sigma'},\label{gpe}
\end{equation}
here the cavity field $\hat a$ in $\hat h_{\sigma\sigma'}$ is replaced by the steady-state cavity amplitude 
\begin{eqnarray}
\alpha=\langle\hat a\rangle=\frac{\eta{\Theta} + \Omega_c{\Xi}}{-U_0{B} + i\kappa},\label{alpha}
\end{eqnarray}
where $B=\langle{\psi}_\downarrow|\cos^2kx|{\psi}_\downarrow\rangle\geq0$ measures the overlap between the optical lattice potential and the density of the spin-$\downarrow$ atoms~\cite{Esslinger10,Nagy08}, ${\Theta}=\langle{\psi}_\downarrow|\cos kx\cos ky|{\psi}_\downarrow\rangle$, and $\Xi= \langle{\psi}_\downarrow|\cos kx\,e^{iky}|{\psi}_\uparrow\rangle$. As will be shown, $\Theta$ and $\Xi$ are two order parameters which determine the configurations of the checkerboard and LVAP phases, respectively. We note that in Eq.~(\ref{alpha}) we have ignored the input noise by assuming that $|U_{0}B|\gg\kappa$, which imposes a minimum pumping strength.

Now, we investigate the quantum phases of the system by numerically finding the steady-state solution of the Eqs. (\ref{gpe}) and (\ref{alpha}). Specifically, we consider a condensate of $N=5\times 10^{3}$ Cr atoms confined in a pancake-shaped trap with $(\nu_{\perp},\nu_{z})=(2\pi)(4,25)\,$kHz. The $s$-wave scattering lengths of the atoms are $a_{\uparrow\uparrow}=a_{\uparrow\downarrow}=a_{\downarrow\uparrow}=112a_{B}$ and $a_{\downarrow\downarrow}=87.5a_{B}$ with $a_{B}$ being the Bohr radius. The wave vector is $k=2\pi/\lambda$ with $\lambda=429.0\,$nm such that the pump laser is blue detuned, $\Delta=-(2\pi)30\,$GHz, with respect to the atomic transition. The single-photon recoil energy is $E_{L}=\hbar^{2}k^{2}/(2M)\simeq  (2\pi)20\,$kHz and the cavity decay rate is taken as $\kappa = 50 E_L$. Furthermore, we assume that the ratio $\Omega_1/g_0$ is fixed at $10$, which implies that $\Omega_{c}=-0.1\Omega$. As a result, the free parameters in our system reduce to the detuning $\delta$, pumping strength $\eta$, and the Raman coupling strength $\Omega$. Finally, for simplicity, we treat the condensate as a quasi-two-dimensional one by assuming that the motion along the $z$ axis is frozen to the ground state of the axial harmonic oscillator.

\begin{figure}[tbp]
\includegraphics[width=0.85\columnwidth]{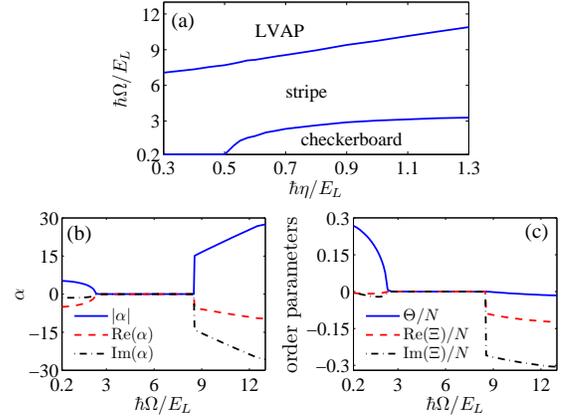}
\caption{(color online). (a) Phase diagram on the $\eta$-$\Omega$ parameter plane with $\delta = -4E_{L}/\hbar$. (b) and (c) show, respectively, the $\Omega$ dependences of the cavity amplitude and order parameters for $\eta=0.7E_{L}/\hbar$.} \label{phase}
\end{figure}

Figure~\ref{phase}(a) summarizes the quantum phases of system on the $\eta$-$\Omega$ parameter plane with $\delta=-4E_{L}/\hbar$. The checkerboard phase originates from the cavity-induced self-organization which occurs for a relatively small $\Omega$ and when the pumping strength exceeds a critical value. For moderate $\Omega$, the system enters the stripe phase as a result of the SO coupling induced by the Raman transition $\Omega$. For even larger $\Omega$, the LVAP phase takes place when the SO coupling induced by $\Omega_{c}$ is comparable to that of $\Omega$ due to the large cavity amplitude. One should note that the phase diagram remains qualitatively unchanged for different $\delta$'s. 

For a fixed pumping strength $\eta=0.7E_{L}/\hbar$, we also plot the cavity amplitude and the order parameters as functions of $\Omega$ in Fig.~\ref{phase}(b) and (c), respectively. As can be seen, $\alpha$ is nonzero in both checkerboard and LVAP phases, indicating that they are superradiant phases. In addition, the fact that $\Theta$ ($\Xi$) is roughly zero if the system is not in the checkerboard (LVAP) phase suggests that the order parameter $\Theta$ ($\Xi$) can indeed be used to characterize the checkerboard (LVAP) phase. In below, we shall discuss the properties of each quantum phases separately.

\begin{figure}[tbp]
\includegraphics[width=0.9\columnwidth]{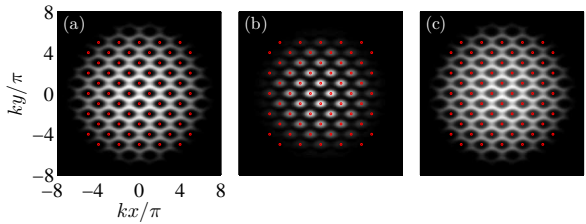}
\caption{(color online). Typical results for the checkerboard phase: (a) $\rho_{\downarrow}(x,y)$, (b) $\rho_{\uparrow}(x,y)$, and (c) $\rho(x,y)$ for $(\delta,\eta,\Omega)=(-4,0.7,1.2)E_{L}/\hbar$. The read dots denote the positions satisfying the condition $\cos(kx)\cos(ky)=-1$.} \label{checkb}
\end{figure}

{\em Checkerboard phase}.---For small $\Omega$, one may neglect the off-diagonal terms in $\hat h$ [Eq. (\ref{singleh})] such that the effective potential for the spin-$\downarrow$ atoms becomes
\begin{align}
M_{z}(x,y)=&\,U_{1}\cos^{2} ky+ U_{0}|\alpha|^{2}\cos^{2} kx\nonumber\\
&+2\eta{\rm Re}(\alpha)\cos kx\cos ky,\label{veffdown}
\end{align}
where ${\rm Re}(\alpha)\approx -\eta U_0 B\Theta/[(U_0B)^2+\kappa^2]$ as $|\Omega_c/\eta| \ll 1$. The first line of Eq. (\ref{veffdown}) represent a 2D $\lambda/2$ periodic potential, where both $U_{1}$ and $U_{0}$ are positive due to the blue-detuned pump laser. As a result, the local minima of $\rho_{\downarrow}=|\psi_{\downarrow}|^{2}$ (henceforth referred to as density holes) locate at the positions satisfying $\cos kx\cos ky=\pm1$. The second line of Eq. (\ref{veffdown}) describes a $\lambda$ periodic potential, under which the density holes of $\rho_{\downarrow}$ locate at either $\cos kx\cos ky=1$ ($\Theta<0$, even sites) or $-1$ ($\Theta>0$, odd sites). The nonlinearity of the system sets in as locations of the density holes depend on the wave function via the order parameter $\Theta$, i.e., the so-called self-organization. The critical pumping strength $\eta^{*}$ for the transition from the $\lambda/2$ to $\lambda$ periodic lattices can be obtain analytically~\cite{Esslinger10}, $\eta^{*}\simeq 0.5E_{L}/\hbar$, in very good agreement with out numerical result.

Figure~\ref{checkb} shows the typical density profiles, $\rho_{\sigma}=|\psi_{\sigma}|^{2}$ and $\rho=\rho_{\uparrow}+\rho_{\downarrow}$, in the checkerboard phase. As can be seen, the density holes on $\rho_{\downarrow}$ match the positions of the odd sites as here we have a positive $\Theta$. On the other hand, $\rho_{\uparrow}$ is mainly determined by the interactions in the two-species condensate. Since the collisional interactions are immiscible, $a_{\uparrow\downarrow}^2>a_{\uparrow\uparrow}a_{\downarrow\downarrow}$, the spin-$\uparrow$ atoms mainly occupy the density holes of the spin-$\downarrow$ atoms. Finally, for the given parameters, the spin-$\downarrow$ atoms dominate in the condensate, therefore, the total density $\rho$ is mainly determined by $\rho_{\downarrow}$.

\begin{figure}[tbp]
\includegraphics[width=0.9\columnwidth]{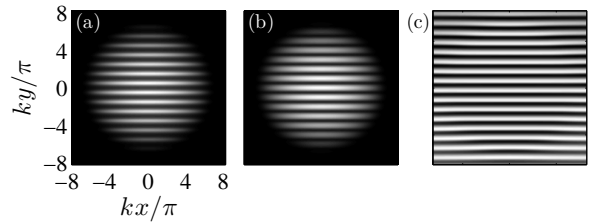}
\caption{(color online). Typical results for the stripe phase: (a) $\rho_{\downarrow}(x,y)$, (b) $\rho_{\uparrow}(x,y)$, (c) $\Delta\phi(x,y)$ for $(\delta,\eta,\Omega)=(-4,1,4.5)E_{L}/\hbar$.} \label{stripep}
\end{figure}

{\em Stripe phase}.---With increasing $\Omega$, one has to take into account the the off-diagonal terms in Eq.~(\ref{singleh}), i.e.,
\begin{eqnarray}
M_{-}(x,y)=(-\Omega \cos ky + \Omega_c \alpha \cos kx)e^{-iky}.\label{veffud}
\end{eqnarray}
Combined with $e^{-iky}$, the first term of $M_{-}$ contains a contribution proportional to $\cos^{2} ky$, which enhances the $\lambda/2$ periodic optical lattice along the $y$ axis in $M_{z}$. In contrast, the periodic potential along the $x$ axis becomes relatively weaker with growing $\Omega$. Consequently, the order parameter gradually $\Theta$ drops to zero, signaling a transition from the checkerboard phase to the stripe phase. Due to the vanishing $\Theta$, the cavity amplitude $\alpha$ also decreases to zero [Fig.~\ref{phase}(b)].

Figure \ref{stripep} shows the typical density profiles and the relative phase of the condensate in the stripe phase. As expected, $\rho_{\downarrow}$ takes the form of a stripe along the $y$ axis with a period $\lambda/2$. Due to the SO coupling, the relative phase of the condensate wave functions $\Delta\phi=\phi_{\uparrow}-\phi_{\downarrow}$ is also periodically modulated along the $y$ axis, where $\phi_{\sigma}={\rm arg}(\psi_{\sigma})$ are the phases of the wave function. To see this more clearly, one may treat the Raman coupling $M_{-}(x,y)$ as an effective transverse magnetic field, i.e., ${\mathbf B}^{\perp}\propto\left[{\rm Re}\left(M_{-}\right),-{\rm Im}\left(M_{-}\right)\right]$. By requiring the pseudo-spin ${\mathbf S}({\mathbf r})=\sum_{\sigma\sigma'} \psi_{\sigma}^*{\hat{\boldsymbol\sigma}}_{\sigma\sigma'}\psi_{\sigma'}$ to be anti-parallel to the direction of the local field, the relative phase can be determined as $\Delta\phi=-2ky$, which is also of period $\lambda/2$. For relatively small $\Omega$, $\rho_{\uparrow}$ is mainly determined by the immiscible interactions [Fig.~\ref{stripep}(b)]. However, for large $\Omega$, the energy associated with the Raman coupling, $E_{\rm Ram}=\int dxdy(\psi_{\uparrow}^{*}M_{-}\psi_{\downarrow}+{\rm c.c})$, becomes important. In fact, it can be shown that the Raman energy, $E_{\rm Ram}=-2\Omega\int dxdy\sqrt{\rho_\uparrow\rho_\downarrow}\cos^2ky$, favors the miscible $\rho_{\downarrow}$ and $\rho_{\uparrow}$, which is also observed in our numerical calculations.

{\em LVAP phase}.---As $\Omega$ is further increased, the cavity induced Raman coupling $\Omega_{c}$ starts to play its role in determining the structure of the system. In fact, as shown in Fig.~\ref{phase}(b) and (c), the mean cavity photon number $|\alpha|$ and the order parameter $\Xi$ simultaneously jump from zero to finite values when $\Omega$ exceeds a critical strength. In addition, the result $|\Xi/\Theta|\gg1$ indicates that a new phase characterized by the order parameter $\Xi$ emerges. 

\begin{figure}[tbp]
\includegraphics[width=0.75\columnwidth]{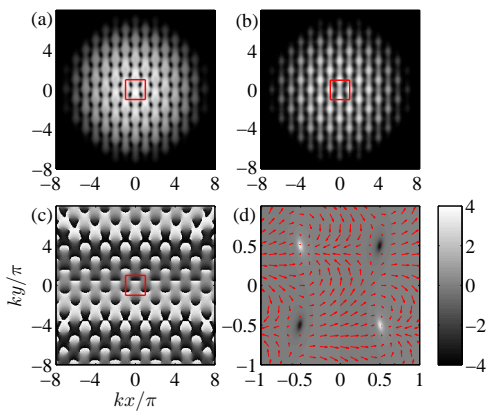}
\caption{(color online). Typical results for the LVAP phase: (a) $\rho_{\downarrow}(x,y)$, (b) $\rho_{\uparrow}(x,y)$, and (c) $\phi_{\downarrow}(x,y)$ for $(\delta,\eta,\Omega)=(-4,0.7,10.35)E_{L}/\hbar$. The red square marks a unit cell. (d) shows ${\mathbf S}_{\perp}$ (vectors) and ${\mathbf B}_{\parallel}$ (grayscale) in a unit cell with the grayscale bar denoting the magnitude of the synthetic magnetic field in units of Tesla.} \label{lvapp}
\end{figure}

Figure~\ref{lvapp}(a) and (b) plot, respectively, $\rho_{\downarrow}$ and $\rho_{\uparrow}$ for the parameters $\hbar(\delta,\eta,\Omega)=(-4,0.7,10.35)E_{L}$. For both spin components, the density holes located at the same positions, $\sin kx\sin ky=1$ (even sites) and $-1$ (odd sites), indicating that for sufficiently large Raman energy the spin-$\uparrow$ and -$\downarrow$ atoms become effectively miscible. Surprisingly, the phase of the wave function $\psi_{\downarrow}$ [Fig.~\ref{lvapp}(c)] reveals that a lattice of vortex-antivortex pairs forms on the spin-$\downarrow$ atoms with vortices (antivortices) located at the odd (even) sites. As shown in Fig.~\ref{lvapp}(a)-(c), if the unit cell is chosen to contain two vortices and two antivortices~\cite{cooper}, the LVAP phase is characterized by lattice constant $\lambda$.

The vector field in Fig.~\ref{lvapp}(d) shows the numerical result of planar spin ${\mathbf S}_{{\perp}}=(S_{x},S_{y})$, which forms a hedgehog (quadrupole) spin texture with winding number $1$ ($-1$) at the odd (even) sites. Since ${\mathbf S}_{{\perp}}$ is anti-parallel to the local transverse field, we expand ${\mathbf B}_{{\perp}}(x,y)$ around the odd or even sites, i.e., ${\mathbf B}_{{\perp}}^{(o)}\propto\left[\Omega_{c}{\rm Im}(\alpha)\varepsilon_x,\Omega \varepsilon_y+\Omega_{c}{\rm Re}(\alpha)\varepsilon_x\right]$ and ${\mathbf B}_{{\perp}}^{(e)}\propto\left[-\Omega_{c}{\rm Im}(\alpha)\varepsilon_{x},\Omega \varepsilon_y-\Omega_{c}{\rm Re}(\alpha)\varepsilon_x\right]$, where $(\varepsilon_{x},\varepsilon_{y})$ is a small deviation from the odd or even sites. As can be seen, for the spin textures to appear, one must have ${\rm Im}(\alpha)\neq0$. Moreover, the positions of hedgehog and quadrupole spin textures are switched by negating $\alpha$. Therefore, the nonlinearity of the LVAP phase embodies in the fact that the positions of the vortices and antivortices on $\psi_{\downarrow}$ depend on the wave function via the order parameter $\Xi$ (here, $\alpha$ is proportional to $\Xi$ when $\Theta$ is negligible).

The cavity field also gives rise to a large synthetic magnetic field in the LVAP phase. To see this, we diagonalize Eq.~(\ref{singleh}) under the adiabatic approximation. Denoting the nondegenerate ground state as $|\chi_{-}\rangle$, the synthetic magnetic field can be expressed as ${\mathbf B}_{{\parallel}}=(i\hbar/e)\nabla\times\langle\chi_{-}|\nabla\chi_{-}\rangle$, which is perpendicular to the $xy$ plane. It can be shown that ${\mathbf B}_{\parallel}$ is nonzero only when $\alpha\neq0$. The grayscal of Fig.~\ref{lvapp}(d) shows ${\mathbf B}_{\parallel}(x,y)$ calculated with the numerically obtained $\alpha$, a structure in analogy to the optical flux lattice~\cite{cooper}. We note that, at exactly the odd and even sites, ${\mathbf B}_{\parallel}$ vanishes due to the lack of Raman coupling at these positions. However, in the vicinity of the odd and even sites, $|{\mathbf B}_{\parallel}|$ quickly reaches a few Tesla, a value much larger than that produced with a spatially dependent magnetic field in Rb condensates ($\sim$mT)~\cite{Lin09}. Since ${\mathbf B}_{\parallel}$ around the odd and even sites point along the opposite directions, the net magnetic flux in each unit cell is zero. Using the same analysis, we find that the similar flux structure can also be found in the setup proposed in Ref.~\cite{xiong}, which allows for the observations of the quantum anomalous Hall effect and chiral topological superfluid phase.

{\em Conclusions and discussions}.---We have proposed a simple scheme to generate SO coupling for a condensate placed inside an optical cavity. Taking Cr condensate as an example, it has been shown that the interplay of the laser lights and the cavity field gives rise to the stripe, checkerboard, and LVAP phases. Particularly, the last two phases stem from the superradiant radiation. We have also shown that, in the LVAP phase, the cavity field induces a nontrivial lattice of flux, which provides a platform for exploring the exotic quantum phases.

The present scheme owns various advantages on the experimental accessibility. The introduction of the optical cavity not only simplifies the laser configuration to a standing wave and a traveling wave (which subsequently reduces the atom heating caused by the extra Raman lasers), but also offers the unique opportunity for monitoring the phase transitions of the system by measuring the cavity output field~\cite{Esslinger10}. Remarkably, in the presence of superradiant radiation, the optical cavity may also provide an additional cooling mechanism for the atoms through the collective emission~\cite{Domokos02, Chan03}. Moreover, unlike the alkali atoms, which suffer from the large heating rate due to spontaneous emission~\cite{Lin11,Wang12,Cheuk12,Zhang12}, the large fine-structure splitting of the proposed Cr atom dramatically suppress the atom heating~\cite{Cui13}. Finally, it is worthwhile to mention that the quantum phases can be detected via the {\it in situ} fluorescence imaging techniques \cite{Chin09,Kuhr10,Reiserer13}.

We are grateful to Han Pu for the helpful discussions. This work was supported by the NSFC (11025421, 11174084, 10935010, and 11121403) and National 973 program (2012CB922104 and 2012CB921904).

\end{document}